\documentclass[twocolumn,aps,pra]{revtex4}
\usepackage[latin1]{inputenc}
\usepackage{units}
\usepackage{amsmath}
\usepackage{amssymb}
\usepackage{stackrel}
\usepackage{esint}

\makeatletter
\@ifundefined{textcolor}{}
{%
 \definecolor{BLACK}{gray}{0}
 \definecolor{WHITE}{gray}{1}
 \definecolor{RED}{rgb}{1,0,0}
 \definecolor{GREEN}{rgb}{0,1,0}
 \definecolor{BLUE}{rgb}{0,0,1}
 \definecolor{CYAN}{cmyk}{1,0,0,0}
 \definecolor{MAGENTA}{cmyk}{0,1,0,0}
 \definecolor{YELLOW}{cmyk}{0,0,1,0}
}

\usepackage{units}
\usepackage{units}
\usepackage[usenames,dvipsnames]{color}
\usepackage{graphicx}
\usepackage{subfigure}
\usepackage{amsfonts}
\usepackage{bm}

\@ifundefined{textcolor}{}{
 \definecolor{BLACK}{gray}{0}
 \definecolor{WHITE}{gray}{1}
 \definecolor{RED}{rgb}{1,0,0}
 \definecolor{GREEN}{rgb}{0,1,0}
 \definecolor{BLUE}{rgb}{0,0,1}
 \definecolor{CYAN}{cmyk}{1,0,0,0}
 \definecolor{MAGENTA}{cmyk}{0,1,0,0}
 \definecolor{YELLOW}{cmyk}{0,0,1,0}
}
\@ifundefined{textcolor}{}{
 \definecolor{BLACK}{gray}{0}
 \definecolor{WHITE}{gray}{1}
 \definecolor{RED}{rgb}{1,0,0}
 \definecolor{GREEN}{rgb}{0,1,0}
 \definecolor{BLUE}{rgb}{0,0,1}
 \definecolor{CYAN}{cmyk}{1,0,0,0}
 \definecolor{MAGENTA}{cmyk}{0,1,0,0}
 \definecolor{YELLOW}{cmyk}{0,0,1,0}
}

\newcommand{\ket}[1]{\ensuremath{\left|{#1}\right\rangle}}

\newcommand{\beq}{\begin{equation}}
\newcommand{\eeq}{\end{equation}}
\newcommand{\bse}{\begin{subequations}}
\newcommand{\ese}{\end{subequations}}
\newcommand{\bea}{\begin{eqnarray}}
\newcommand{\eea}{\end{eqnarray}}

\makeatother

\begin{document}

\title{On the spin projection operator and the probabilistic meaning\\
 of the bipartite correlation function}

\author{Ana María Cetto, Andrea Valdés-Hernández and Luis de la Peña}

\affiliation{Instituto de Física, Universidad Nacional Autónoma de México, A.
P. 20-364, Ciudad de México, Mexico}
\begin{abstract}
Spin is a fundamental and distinctive property of the electron, having
far-reaching consequences in wide areas of physics. Yet, further to
its association with an angular momentum, the physics underpinning
its formal treatment remains obscure. In this work we propose to advance
in disclosing the meaning behind the formalism, by first recalling
some basic facts about the one-particle spin operator. Consistently
informed by and in line with the quantum formalism, we then proceed
to analyse in detail the spin projection operator correlation function
$C_{Q}(\boldsymbol{a},\boldsymbol{b})=\left\langle \left(\hat{\boldsymbol{\sigma}}\cdotp\boldsymbol{a}\right)\left(\hat{\boldsymbol{\sigma}}\cdotp\boldsymbol{b}\right)\right\rangle $
for the bipartite singlet state, and show it to be amenable to an
unequivocal probabilistic reading. In particular, the calculation
of $C_{Q}(\boldsymbol{a},\boldsymbol{b})$ entails a partitioning
of the probability space, which is dependent on the directions $(\boldsymbol{a},\boldsymbol{b})$.
The derivation of the CHSH- or other Bell-type inequalities, on the
other hand, does not consider such partitioning. This observation
puts into question the applicability of Bell-type inequalities to
the bipartite singlet spin state. 
\end{abstract}
\maketitle

\section{Introduction}

Ever since Pauli's dictum on the impossibility of a model for the
electron spin, physicists have been taught to replace it with the
abstract concept of spin as a dichotomic quantity living in its own
Hilbert space \cite{hist}. The introduction of spin as a postulate
in nonrelativistic quantum mechanics and its successful treatment
in terms of Pauli matrices and spinors seem to foreclose the need
for a deeper reflection on its nature. And yet the very existence
of the electron spin has a huge impact in various areas of physics,
ranging from atomic structure to fermion statistics, the stability
of matter, and quantum communication.

This paper is devoted to a close analysis of the spin operator and
associated observables, with the intention to advance in the understanding
of their meaning. The analysis carried out, fully within the conventional
quantum formalism, serves to clarify certain assumptions usually made
regarding the correspondence between the mathematical expressions
involving spin operators and their geometrical interpretation. Such
clarification bears particular relevance in the case of the (entangled)
singlet state of the two-electron system, putting into question the
interpretation of the Bell-type inequalities as conventionally applied
to this case.

The paper is organized as follows. In section II, basic elements of
the quantum description of the single spin $\nicefrac{1}{2}$ are
reviewed. Section III contains a brief introduction to the entangled
bipartite system, in preparation for the discussion on the physical
meaning of the spin projection operator correlation function $C_{Q}(\boldsymbol{a},\boldsymbol{b})$
in section IV, where an appropiate disaggregation of $C_{Q}(\boldsymbol{a},\boldsymbol{b})$
in terms of the individual spin projection eigenfunctions is carried
out, leading to expressions with a clear probabilistic meaning. In
Section V these expressions are translated to the hidden-variable
language, in order to make contact with the Bell-type inequalities.
The paper concludes with a discussion on the applicability of such
inequalities to the bipartite entangled spin state, and relates this
discussion to relevant literature on the subject.

\section{Single-spin state and spin projections}

Let us start by recalling that the most general one-particle (pure)
spin $\nicefrac{1}{2}-$state can be expressed in terms of the spinor
(except for an irrelevant overall factor $e^{i\eta}$) 
\begin{equation}
\left|+_{r}\right\rangle =\cos\frac{\theta}{2}\left|+_{z}\right\rangle +e^{i\varphi}\sin\frac{\theta}{2}\left|-_{z}\right\rangle ,\label{2-2}
\end{equation}
and the spinor orthogonal to it, 
\begin{equation}
\left|-_{r}\right\rangle =-e^{-i\varphi}\sin\frac{\theta}{2}\left|+_{z}\right\rangle +\cos\frac{\theta}{2}\left|-_{z}\right\rangle ,\label{2-3}
\end{equation}
with $0\leq\theta\leq\pi$ and $0\leq\varphi\leq2\pi$, $\theta$
and $\varphi$ being the zenithal and azimuthal angles that define
the unit vector 
\begin{equation}
\boldsymbol{r}=\mathbf{i}\sin\theta\cos\varphi+\mathbf{j}\sin\theta\sin\varphi+\mathbf{k}\cos\theta\label{2-4}
\end{equation}
in 3D space, identified in the literature as the Bloch vector (see,
e. g., Ref. \cite{JA}).

In terms of the Pauli matrices $\hat{\sigma}_{i}$ ($i=x,y,z)$, the
spin operator is given by $\mathbf{\hat{s}}=(\hbar/2)\mathbf{\hat{\boldsymbol{\sigma}}}$;
however, we shall refer throughout to $\mathbf{\hat{\boldsymbol{\sigma}}}$
as the spin operator, for simplicity. The states $\left|\pm_{z}\right\rangle $
are such that 
\begin{equation}
\hat{\boldsymbol{\sigma}}\cdotp\mathbf{k}\left|\pm_{z}\right\rangle =\pm\left|\pm_{z}\right\rangle ,\label{sigz}
\end{equation}
i. e., they are eigenstates of the spin projection operator along
$\mathbf{k}$. The vectors $\left|\pm_{r}\right\rangle $, in their
turn, which can be obtained by applying the unitary rotation matrix
\begin{equation}
U(\theta,\varphi)=\left(\begin{array}{cc}
\cos\frac{\theta}{2} & -e^{-i\varphi}\sin\frac{\theta}{2}\\
e^{i\varphi}\sin\frac{\theta}{2} & \cos\frac{\theta}{2}
\end{array}\right)\label{2-6-1}
\end{equation}
to the spinors 
\begin{equation}
\left|+_{z}\right\rangle =\left(\begin{array}{cc}
1\\
0
\end{array}\right),\;\left|-_{z}\right\rangle =\left(\begin{array}{cc}
0\\
1
\end{array}\right),\label{2-6-2}
\end{equation}
satisfy the eigenvalue equation 
\begin{equation}
\hat{\boldsymbol{\sigma}}\cdotp\boldsymbol{r}\left|\pm_{r}\right\rangle =\pm\left|\pm_{r}\right\rangle .\label{2-4'}
\end{equation}
Therefore, $\left|+_{r}\right\rangle $ and $\left|-_{r}\right\rangle $
correspond to states representing, respectively, parallel and antiparallel
spin projections along the arbitrary direction $\boldsymbol{r}$.
The expectation value of $\boldsymbol{\hat{\sigma}}$ in the states
(\ref{2-2}) and (\ref{2-3}) is 
\begin{equation}
\left\langle +_{r}\right|\boldsymbol{\hat{\sigma}}\left|+_{r}\right\rangle =\boldsymbol{r},\quad\left\langle -_{r}\right|\boldsymbol{\hat{\sigma}}\left|-_{r}\right\rangle =-\boldsymbol{r}\label{2'-5}
\end{equation}
whence one may associate indeed the direction $\boldsymbol{r}$ with
the spin state $\left|+_{r}\right\rangle $, and similarly the direction
$-\boldsymbol{r}$ with the spin state $\left|-_{r}\right\rangle $.

Consider now a second arbitrary direction, determined by the unitary
vector $\boldsymbol{a}=(a_{x},a_{y},a_{z})$. The projection of the
spin operator along this new direction takes on the matrix form 
\begin{equation}
\hat{\boldsymbol{\sigma}}\cdotp\boldsymbol{a}=\left(\begin{array}{cc}
a_{z} & a_{x}-ia_{y}\\
a_{x}+ia_{y} & -a_{z}
\end{array}\right),\label{2-3'-1}
\end{equation}
and its expectation values read 
\begin{equation}
\left\langle \pm_{r}\right|\hat{\boldsymbol{\sigma}}\cdotp\boldsymbol{a}\left|\pm_{r}\right\rangle =\pm\boldsymbol{r}\cdot\boldsymbol{a}=\pm\cos\theta_{ra},\label{2-4b}
\end{equation}
with $\theta_{ra}$ the angle formed by $\boldsymbol{r}$ and $\boldsymbol{a}$.
In their turn, the off-diagonal elements of $\boldsymbol{\hat{\sigma}}\cdot\boldsymbol{a}$
in the basis $\{\left|+_{r}\right\rangle ,\left|-_{r}\right\rangle \}$
are given by 
\begin{eqnarray}
\left\langle -_{r}\right|\hat{\boldsymbol{\sigma}}\cdotp\boldsymbol{a}\left|+_{r}\right\rangle  & = & \left\langle +_{r}\right|\hat{\boldsymbol{\sigma}}\cdotp\boldsymbol{a}\left|-_{r}\right\rangle ^{*}\nonumber \\
 & = & e^{i\varphi}(\boldsymbol{\theta}+i\boldsymbol{\varphi})\cdot\boldsymbol{a},\label{2-4c}
\end{eqnarray}
where $\boldsymbol{\theta}$ and $\boldsymbol{\varphi}$ are the unit
vectors that, together with $\boldsymbol{r}$ given by (\ref{2-4}),
define the orthogonal triad in spherical coordinates, 
\begin{eqnarray}
\boldsymbol{\theta} & = & \mathbf{i}\cos\theta\cos\varphi+\mathbf{j}\cos\theta\sin\varphi-\mathbf{k}\sin\theta,\nonumber \\
\boldsymbol{\varphi} & = & -\mathbf{i}\sin\theta+\mathbf{j}\cos\varphi.
\end{eqnarray}
Further, the modulus of (\ref{2-4c}) is given by
\begin{equation}
\left|\left\langle -_{r}\right|\hat{\boldsymbol{\sigma}}\cdotp\boldsymbol{a}\left|+_{r}\right\rangle \right|=\mid\boldsymbol{r}\times\boldsymbol{a}\mid=\sin\theta_{ra},\label{4-4}
\end{equation}
which should come as no surprise, since 
\begin{align}
1 & =\left\langle +_{r}\right|\left(\hat{\boldsymbol{\sigma}}\cdotp\boldsymbol{a}\right)^{2}\left|+_{r}\right\rangle \nonumber \\
 & =\left\langle +_{r}\right|\left(\hat{\boldsymbol{\sigma}}\cdotp\boldsymbol{a}\right)\big[\left|+_{r}\right\rangle \left\langle +_{r}\right|+\left|-_{r}\right\rangle \left\langle -_{r}\right|\big]\left(\hat{\boldsymbol{\sigma}}\cdotp\boldsymbol{a}\right)\left|+_{r}\right\rangle ,\label{4-6}
\end{align}
whence, using (\ref{2-4b}), one obtains $\left|\left\langle -_{r}\right|\hat{\boldsymbol{\sigma}}\cdotp\boldsymbol{a}\left|+_{r}\right\rangle \right|{}^{2}=1-\cos\theta_{ra}^{2}=\sin^{2}\theta_{ra}$.

It is interesting to note that the operator $\boldsymbol{\hat{\sigma}}\cdot\boldsymbol{a}$,
rather than simply projecting the spin vector onto $\boldsymbol{a}$,
changes in a nontrivial way the direction associated with it. In terms
of the angles ($\theta_{a},\varphi_{a}$) that define the orientation
of $\boldsymbol{a}$, we have
\begin{equation}
\left(\hat{\mathbf{\boldsymbol{\sigma}}}\cdotp\boldsymbol{a}\right)\left|+_{r}\right\rangle =\left(\begin{array}{c}
\cos\theta_{a}\cos\frac{\theta}{2}+\sin\theta_{a}e^{i\left(\varphi-\varphi_{a}\right)}\sin\frac{\theta}{2}\\
\sin\theta_{a}e^{i\varphi_{a}}\cos\frac{\theta}{2}-\cos\theta_{a}e^{i\varphi}\sin\frac{\theta}{2}
\end{array}\right).\label{2-8}
\end{equation}
This vector, along with 
\begin{equation}
\left(\hat{\mathbf{\boldsymbol{\sigma}}}\cdotp\boldsymbol{a}\right)\left|-_{r}\right\rangle =\left(\begin{array}{c}
-\cos\theta_{a}e^{-i\varphi}\sin\frac{\theta}{2}+\sin\theta_{a}e^{-i\varphi_{a}}\cos\frac{\theta}{2}\\
-\sin\theta_{a}e^{-i\left(\varphi-\varphi_{a}\right)}\sin\frac{\theta}{2}-\cos\theta_{a}\cos\frac{\theta}{2}
\end{array}\right),\label{2-8'}
\end{equation}
form an orthonormal basis, with a different spin orientation in 3D
space (except of course when $\boldsymbol{a}$ and $\boldsymbol{r}$
are colinear, as shown in (\ref{2-4'})). The corresponding Bloch
vector, which we call $\boldsymbol{r_{a}}$, is obtained by taking
the expectation value
\begin{equation}
\left\langle +_{ra}\right|\boldsymbol{\hat{\sigma}}\left|+_{ra}\right\rangle \boldsymbol{=r_{a}},\label{2-9}
\end{equation}
with
\begin{equation}
\left|+_{ra}\right\rangle \equiv\left(\hat{\mathbf{\boldsymbol{\sigma}}}\cdotp\boldsymbol{a}\right)\left|+_{r}\right\rangle .\label{2-9'}
\end{equation}
The calculation of (\ref{2-9}) is most easily done by using once
more $\left|+_{r}\right\rangle \left\langle +_{r}\right|+\left|-_{r}\right\rangle \left\langle -_{r}\right|=\mathbb{I}$,
together with Eqs. (\ref{2-4b}) and (\ref{2-4c}); the result,
\begin{equation}
\boldsymbol{r_{a}}=2\left(\boldsymbol{r}\cdot\boldsymbol{a}\right)\boldsymbol{a}-\boldsymbol{r},\label{2-9''}
\end{equation}
clearly depends on both vectors $\boldsymbol{r}$ and $\boldsymbol{a}$,
not just on the angle formed by them. Interestingly, a second operation
$\left(\hat{\mathbf{\boldsymbol{\sigma}}}\cdotp\boldsymbol{a}\right)$
performed on $\left|+_{r}\right\rangle $ (or on any spin state vector,
for that matter) brings it back to its original form, or in terms
of (\ref{2-9'}),
\begin{equation}
\left(\hat{\mathbf{\boldsymbol{\sigma}}}\cdotp\boldsymbol{a}\right)\left|+_{ra}\right\rangle =\left|+_{r}\right\rangle ,\label{2-10'}
\end{equation}
since $\left(\hat{\mathbf{\boldsymbol{\sigma}}}\cdotp\boldsymbol{a}\right)\left(\hat{\mathbf{\boldsymbol{\sigma}}}\cdotp\boldsymbol{a}\right)=1$.
In other words, the effect of $\left(\hat{\mathbf{\boldsymbol{\sigma}}}\cdotp\boldsymbol{a}\right)$
on a state vector is reversible. \cite{Footnote1}

\section{Bipartite spin state and associated observables\label{bp}}

Let us now consider a system of two $\nicefrac{1}{2}-$spin particles
in the (entangled) singlet state 
\begin{equation}
\left\vert \Psi^{0}\right\rangle =\frac{1}{\sqrt{2}}\left(\left\vert +_{r}\right\rangle \left\vert -_{r}\right\rangle -\left\vert -_{r}\right\rangle \left\vert +_{r}\right\rangle \right),\label{3-2}
\end{equation}
in terms of the simplified (standard) notation $\left\vert {\phi}\right\rangle \left\vert {\chi}\right\rangle =\left\vert {\phi}\right\rangle \otimes\left\vert {\chi}\right\rangle ,$
with $\left\vert {\phi}\right\rangle $ a vector in the Hilbert space
of subsystem 1, and $\left\vert {\chi}\right\rangle $ a vector in
the Hilbert space of subsystem 2. It is convenient to bear in mind
that the angles $(\theta,\varphi)$ that define the direction of $\boldsymbol{r}$
are not set; the singlet state is spherically symmetric, and therefore
$\boldsymbol{r}$ may be chosen arbitrarily. 

Now, in the composite system, the projection of the first spin operator
along an arbitrary direction $\boldsymbol{a}$ is described by $(\hat{\boldsymbol{\sigma}}\cdotp\boldsymbol{a})\otimes\mathbb{I}$,
and the projection of the second spin operator along a direction $\boldsymbol{b}$
is described by $\mathbb{I}\otimes(\hat{\boldsymbol{\sigma}}\cdotp\boldsymbol{b})$.
The expectation value of any of these operators in the state $\left|{\Psi^{0}}\right\rangle $
vanishes, 
\begin{eqnarray}
\left\langle \Psi^{0}\right|\left(\hat{\boldsymbol{\sigma}}\cdotp\boldsymbol{a}\right)\otimes\mathbb{I}\left|\Psi^{0}\right\rangle  & = & \frac{1}{2}\big(\left\langle +_{r}\right|\hat{\boldsymbol{\sigma}}\cdotp\boldsymbol{a}\left|+_{r}\right\rangle +\left\langle -_{r}\right|\hat{\boldsymbol{\sigma}}\cdotp\boldsymbol{a}\left|-_{r}\right\rangle \big)\nonumber \\
 & = & 0,\label{3-4}
\end{eqnarray}
as follows from Eq. (\ref{2-4b}). Further, their quantum correlation
\begin{equation}
C_{Q}(\boldsymbol{a},\boldsymbol{b})=\left\langle \Psi^{0}\right|\left(\hat{\boldsymbol{\sigma}}\cdotp\boldsymbol{a}\right)\otimes\left(\hat{\boldsymbol{\sigma}}\cdotp\boldsymbol{b}\right)\left|\Psi^{0}\right\rangle \label{CQab}
\end{equation}
is given according to Eqs. (\ref{2-4b}) and (\ref{2-4c}) by 
\begin{gather}
C_{Q}(\boldsymbol{a},\boldsymbol{b})=\nonumber \\
=-\big[(\boldsymbol{r}\cdot\boldsymbol{a})(\boldsymbol{r}\cdot\boldsymbol{b})+(\boldsymbol{\theta}\cdot\boldsymbol{a})(\boldsymbol{\theta}\cdot\boldsymbol{b})+(\boldsymbol{\varphi}\cdot\boldsymbol{a})(\boldsymbol{\varphi}\cdot\boldsymbol{b})\big]\nonumber \\
=-\boldsymbol{a}\cdot\boldsymbol{b}.\label{CQ}
\end{gather}
Though this result is well known, it is worthwhile revisiting it from
a different angle, as follows.

We take as an orthonormal and complete basis of the composite Hilbert
space, the set of vectors 
\begin{eqnarray}
\left|\Psi^{1}\right\rangle  & = & \left|+_{r}\right\rangle \left|-_{r}\right\rangle ,\;\;\left|\Psi^{2}\right\rangle =\left|-_{r}\right\rangle \left|+_{r}\right\rangle ,\nonumber \\
\left|\Psi^{3}\right\rangle  & = & \left|+_{r}\right\rangle \left|+_{r}\right\rangle ,\;\;\left|\Psi^{4}\right\rangle =\left|-_{r}\right\rangle \left|-_{r}\right\rangle ,\label{3-6}
\end{eqnarray}
and write 
\begin{eqnarray}
C_{Q}(\boldsymbol{a},\boldsymbol{b}) & = & \left\langle \Psi^{0}\right|\left(\hat{\boldsymbol{\sigma}}\cdotp\boldsymbol{a}\right)\Big(\sum_{k=1}^{4}\left|\Psi^{k}\right\rangle \left\langle \Psi^{k}\right|\Big)\left(\hat{\boldsymbol{\sigma}}\cdotp\boldsymbol{b}\right)\left|\Psi^{0}\right\rangle \nonumber \\
 & = & \sum_{k=1}^{4}\left\langle \Psi^{0}\right|(\hat{\boldsymbol{\sigma}}\cdotp\boldsymbol{a})\otimes\mathbb{I}\left|\Psi^{k}\right\rangle \left\langle \Psi^{k}\right|\mathbb{I}\otimes(\hat{\boldsymbol{\sigma}}\cdotp\boldsymbol{b})\left|\Psi^{0}\right\rangle \nonumber \\
 & = & \sum_{k=1}^{4}F_{k}.\label{3-8}
\end{eqnarray}
Calculation of the separate terms that contribute to $C_{Q}(\boldsymbol{a},\boldsymbol{b})$
gives, with the aid of Eqs. (\ref{2-4b}) and (\ref{2-4c}), 
\begin{eqnarray}
F_{1} & = & \frac{1}{2}\left\langle +_{r}\right|\hat{\boldsymbol{\sigma}}\cdotp\boldsymbol{a}\left|+_{r}\right\rangle \left\langle -_{r}\right|\hat{\boldsymbol{\sigma}}\cdotp\boldsymbol{b}\left|-_{r}\right\rangle \nonumber \\
 & = & -\frac{1}{2}(\boldsymbol{r}\cdot\boldsymbol{a})(\boldsymbol{r}\cdot\boldsymbol{b}),\\
F_{2} & = & \frac{1}{2}\left\langle -_{r}\right|\hat{\boldsymbol{\sigma}}\cdotp\boldsymbol{a}\left|-_{r}\right\rangle \left\langle +_{r}\right|\hat{\boldsymbol{\sigma}}\cdotp\boldsymbol{b}\left|+_{r}\right\rangle =F_{1},\\
F_{3} & = & -\frac{1}{2}\left\langle -_{r}\right|\hat{\boldsymbol{\sigma}}\cdotp\boldsymbol{a}\left|+_{r}\right\rangle \left\langle +_{r}\right|\hat{\boldsymbol{\sigma}}\cdotp\boldsymbol{b}\left|-_{r}\right\rangle \nonumber \\
 & = & -\frac{1}{2}\left[(\boldsymbol{r}\times\boldsymbol{a})\cdot(\boldsymbol{r}\times\boldsymbol{b})-i\boldsymbol{r}\cdot(\boldsymbol{a}\times\boldsymbol{b})\right],\\
F_{4} & = & -\frac{1}{2}\left\langle +_{r}\right|\hat{\boldsymbol{\sigma}}\cdotp\boldsymbol{a}\left|-_{r}\right\rangle \left\langle -_{r}\right|\hat{\boldsymbol{\sigma}}\cdotp\boldsymbol{b}\left|+_{r}\right\rangle =F_{3}^{*}.
\end{eqnarray}
The sum of the four terms gives of course the expected expression
(\ref{CQ}). The fact that this result depends only on the angle formed
by $\boldsymbol{a}$ and $\boldsymbol{b}$ is due to the spherical
symmetry of the singlet spin state. It is interesting however to look
at the terms separately, and observe that the sum of the first two
($F_{1}+F_{2}$), which involve intermediate states ($\left|{\Psi^{1}}\right\rangle $
and $\left|{\Psi^{2}}\right\rangle $) of \emph{antiparallel} spins
(along the arbitrary direction $\boldsymbol{r}$), gives the product
of the projections of $\boldsymbol{a}$ and $\boldsymbol{b}$ onto
$\boldsymbol{r}$, whilst the sum of the last two ($F_{3}+F_{4}$),
which involve intermediate states ($\left|{\Psi^{3}}\right\rangle $
and $\left|{\Psi^{4}}\right\rangle $) of \emph{parallel} spins, contains
their vector products. In other words, the two spin projection operators
$\hat{\boldsymbol{\sigma}}\cdotp\boldsymbol{a}$, $\hat{\boldsymbol{\sigma}}\cdotp\boldsymbol{b}$
establish a correlation not just through the intermediate states representing
\emph{antiparallel} spins\textemdash as one might suppose for the
entangled spin-zero state\textemdash but also through the intermediate
states of \emph{parallel} spins, $\left|{+_{r}}\right\rangle \left|{+_{r}}\right\rangle $
and $\left|{-_{r}}\right\rangle \left|{-_{r}}\right\rangle $. \cite{Footnote2}

\section{Probabilistic content of the spin projection correlation function}

With the above elements we have prepared the ground to address the
question: what exactly is the physical content of Eq. (\ref{CQab})?
It is commonplace to say simply that it is the average of a product
of spin projections, and this is what is put experimentally to test.
Here we hope to contribute to provide a fairer picture of this expression
by analysing it in more detail.

We focus again on the quantum correlation for the bipartite singlet
spin state as expressed in Eq. (\ref{CQab}), and make an alternative
calculation, this time resorting to the eigenvalue equations 
\begin{align}
 & \hat{\boldsymbol{\sigma}}\cdotp\boldsymbol{a}\left|{\pm_{a}}\right\rangle =\alpha\left|{\pm_{a}}\right\rangle ,\ \alpha=\pm1,\nonumber \\
 & \hat{\boldsymbol{\sigma}}\cdotp\boldsymbol{b}\left|{\pm_{b}}\right\rangle =\beta\left|{\pm_{b}}\right\rangle ,\ \beta=\pm1,\label{5-2-1}
\end{align}
to construct a new orthonormal basis for the bipartite system: 
\[
\left|{\phi^{1}}\right\rangle =\left|{+_{a}}\right\rangle \left|{-_{b}}\right\rangle ,\ \ \left|{\phi^{2}}\right\rangle =\left|{-_{a}}\right\rangle \left|{+_{b}}\right\rangle ,
\]
\begin{equation}
\left|{\phi^{3}}\right\rangle =\left|{+_{a}}\right\rangle \left|{+_{b}}\right\rangle ,\ \ \left|{\phi^{4}}\right\rangle =\left|{-_{a}}\right\rangle \left|{-_{b}}\right\rangle .\label{5-6-1}
\end{equation}
Instead of (\ref{3-8}) we then write 
\begin{equation}
C_{Q}(\boldsymbol{a},\boldsymbol{b})=\left\langle \Psi^{0}\right|(\hat{\boldsymbol{\sigma}}\cdotp\boldsymbol{a})\left(\sum_{k=1}^{4}\left|\phi^{k}\right\rangle \left\langle \phi^{k}\right|\right)(\hat{\boldsymbol{\sigma}}\cdotp\boldsymbol{b})\left|\Psi^{0}\right\rangle .\label{5-4-1}
\end{equation}
The basis $\left\{ \left|\phi^{k}\right\rangle \right\} $ is a very
convenient one, given that, in view of (\ref{5-2-1}) and (\ref{5-6-1}),
\begin{equation}
(\hat{\boldsymbol{\sigma}}\cdotp\boldsymbol{a})\otimes\mathbb{I}\left|\phi^{k}\right\rangle \left\langle \phi^{k}\right|\mathbb{I}\otimes(\hat{\boldsymbol{\sigma}}\cdotp\boldsymbol{b})=A_{k}\left|\phi^{k}\right\rangle \left\langle \phi^{k}\right|,\label{5-9}
\end{equation}
with 
\begin{equation}
A_{k}=\alpha_{k}\beta{}_{k},\label{5-7}
\end{equation}
where $\alpha_{k},\beta_{k}$ are the individual eigenvalues corresponding
to the bipartite state $\left|\phi^{k}\right\rangle $. Thus from
Eq. (\ref{5-9}) we get 
\begin{equation}
C_{Q}(\boldsymbol{a},\boldsymbol{b})=\sum_{k}A_{k}(\boldsymbol{a},\boldsymbol{b})C_{k}(\boldsymbol{a},\boldsymbol{b}),\label{5-10}
\end{equation}
with 
\begin{equation}
C_{k}=|\langle\phi^{k}|\Psi^{0}\rangle|^{2}.\label{5-11}
\end{equation}
From Eqs. (\ref{5-2-1}-\ref{5-7}) it follows that 
\begin{equation}
A_{1}=A_{2}=-1,\ A_{3}=A_{4}=+1.\label{5-9'}
\end{equation}

Note that the coefficients in Eq. (\ref{5-10}) have an unambiguous
meaning: $A_{k}$ is the eigenvalue of the operator $\left(\hat{\boldsymbol{\sigma}}\cdotp\boldsymbol{a}\otimes\hat{\boldsymbol{\sigma}}\cdotp\boldsymbol{b}\right)$
corresponding to the bipartite state $\left\vert \phi^{k}\right\rangle $.
In its turn, $C_{k}$ is the relative weight of the eigenvalue $A_{k}$.
Further, the $C_{k}$ are nonnegative and add to give
\begin{equation}
\sum_{k}C_{k}=\sum_{k}\langle\Psi^{0}|\phi^{k}\rangle\langle\phi^{k}|\Psi^{0}\rangle=1.\label{5-13}
\end{equation}
Consequently, $C_{k}$ can be identified with the joint probability
associated with the corresponding $A_{k}$, 
\begin{equation}
C_{k}(\boldsymbol{a},\boldsymbol{b})=P_{ab}^{k}(\alpha,\beta),\label{5-14}
\end{equation}
or in explicit terms, using (\ref{5-6-1}) (the superindex $k$ is
now redundant), 
\[
C_{1}(\boldsymbol{a},\boldsymbol{b})=P_{ab}(+,-),\ C_{2}(\boldsymbol{a},\boldsymbol{b})=P_{ab}(-,+),
\]
\begin{equation}
C_{3}(\boldsymbol{a},\boldsymbol{b})=P_{ab}(+,+),\ C_{4}(\boldsymbol{a},\boldsymbol{b})=P_{ab}(-,-).\label{5-15}
\end{equation}
The marginal probability $P_{ab}(\alpha)$ is obtained from these
expressions by considering the two possible outcomes $\beta=\pm1$
for a given $\alpha$, so for instance (using (\ref{5-11}))
\begin{eqnarray}
P_{ab}(\alpha=+) & = & P_{ab}(+,-)+P_{ab}(+,+)\nonumber \\
 & = & \langle\phi^{1}|\Psi^{0}\rangle\langle\Psi^{0}|\phi^{1}\rangle+\langle\phi^{3}|\Psi^{0}\rangle\langle\Psi^{0}|\phi^{3}\rangle\nonumber \\
 & = & \langle+_{a}|\big[\langle-_{b}|\Psi^{0}\rangle\langle\Psi^{0}|-_{b}\rangle\nonumber \\
 &  & +\langle+_{b}|\Psi^{0}\rangle\langle\Psi^{0}|+_{b}\rangle\big]|+_{a}\rangle\nonumber \\
 & = & \langle+_{a}|\mathrm{Tr_{2}}\big(|\Psi^{0}\rangle\langle\Psi^{0}|\big)|+_{a}\rangle\\
 & = & \frac{1}{2},\label{5-16}
\end{eqnarray}
where $\textrm{Tr}_{2}\big(|\Psi^{0}\rangle\langle\Psi^{0}|\big)$
denotes the partial trace of $|\Psi^{0}\rangle\langle\Psi^{0}|$ over
the degrees of freedom of subsystem $2$, and hence represents the
(reduced) density matrix of the first subsystem, $\rho_{1}$. Since
$\ket{\Psi^{0}}$ is a maximally entangled state, $\rho_{1}=(1/2)\mathbb{I}$,
which leads directly to the result (\ref{5-16}). The same applies
of course to all four marginal probabilities.

The joint probability (\ref{5-14}) can also be written as the product
of the marginal probability of occurrence of a given $\beta$ and
the conditional probability $P_{ab}(\alpha\mid\beta)$. Thus for instance,
\[
P_{ab}(+,-)=P_{ab}(+\mid-)P_{ab}(\beta=-),
\]
with $P_{ab}(+\mid-)$ the probability of occurrence of $\alpha=+$
under the condition that $\beta=-1$. A further interesting result
in support of the probabilistic meaning just described, is obtained
by integrating $P_{ab}(\alpha,\beta)$ over all possible orientations
of $\boldsymbol{b}$ to get the probability of $\alpha$ having a
given value (say, $\alpha=1)$, with $\beta=1$ in any direction ($\Omega_{b}$
is the solid angle), 
\[
\int\mathrm{d\Omega_{b}}P_{ab}(+,+)=\frac{1}{4}.
\]

Since the basis $\left\{ \left\vert \phi^{k}\right\rangle \right\} $
was constructed in terms of the individual eigenvectors $\left\vert {\pm_{a}}\right\rangle $
and $\left\vert {\pm_{b}}\right\rangle $, it is essential to use
this basis \emph{consistently} in the calculations leading to $C_{Q}(\boldsymbol{a},\boldsymbol{b})$;
in other words, both the relative weights (or joint probabilities)
$C_{k}$ and the eigenvalues $A_{k}$ are anchored to this basis.
To stress this point, in Eq. (\ref{5-10}) the dependence on $\boldsymbol{a}$
and $\boldsymbol{b}$ has been introduced explicitly in the notation.
If a different direction $\boldsymbol{b}^{\prime}$ is chosen for
the calculation of $C_{Q}(\boldsymbol{a},\boldsymbol{b}^{\prime})$,
a new vector basis $\{\left\vert \phi^{\prime k}\right\rangle \}$
will have to be used, with elements involving the eigenstates of the
operator $\hat{\boldsymbol{\sigma}}\cdotp\boldsymbol{b}^{\prime}$.
Clearly this will lead in general to different values for the individual
coefficients $C_{k}(\boldsymbol{a},\boldsymbol{b}^{\prime})$.

\section{Discussion}

A corollary of the anaysis carried out above is that the partition
of the probability space $\Lambda$ into subspaces appropriate for
the construction of $C_{Q}(\boldsymbol{a},\boldsymbol{b})$, cannot
be the same as that used to construct $C_{Q}(\boldsymbol{a},\boldsymbol{b}^{\prime})$.
This important restriction, which here emerges as a direct consequence
of the operator algebra, is nevertheless often overlooked or not well
understood.

To clarify this point, let us go back to Eq. (\ref{5-10}) and note
that by using the basis constructed with the eigenvectors of the individual
(commuting) spin projection operators $\left(\hat{\boldsymbol{\sigma}}\cdotp\boldsymbol{a}\right)$,
$\left(\hat{\boldsymbol{\sigma}}\cdotp\boldsymbol{b}\right)$, pertaining
to particles 1 and 2, respectively, we have been able to write $C_{Q}(\boldsymbol{a},\boldsymbol{b})$
as a sum of (product) eigenvalues $A_{k}$ with their corresponding
statistical weights $C_{k}$. This means that the ensemble of systems
represented by the entangled state vector $\left|{\Psi^{0}}\right\rangle $,
has been partitioned into four subensembles that are mutually exclusive
and complementary: the subensembles that produce the outcomes $(+,-)$,
$(-,+)$, $(+,+)$ and $(-,-)$ for $(\alpha,\beta)$, given a certain
pair of directions $(\boldsymbol{a},\boldsymbol{b})$. These subensembles
are represented by the basis vectors $\left|\phi^{k}\right\rangle $,
$k=1,2,3,4$. Every (bipartite) element of the full ensemble belongs
to one and only one of such subensembles. For a different pair $(\boldsymbol{a},\boldsymbol{b}^{\prime})$,
the partitioning of the (same) ensemble of systems represented by
$\left|{\Psi^{0}}\right\rangle $ will be into four (mutually exclusive
and complementary) subensembles \emph{different} from the previous
ones: those that produce the outcomes $(+,-^{\prime})$, $(-,+^{\prime})$,
$(+,+^{\prime})$ and $(-,-^{\prime})$ for a pair of directions $(\boldsymbol{a},\boldsymbol{b}^{\prime})$,
and are represented by the basis vectors $\left|\phi^{\prime k}\right\rangle $
.

It is clear from this discussion that an expression that combines
eigenvalues $A_{k}$, $A_{k}^{\prime}$ pertaining to different pairs
$(\boldsymbol{a},\boldsymbol{b})$, $(\boldsymbol{a},\boldsymbol{b}^{\prime})$
is physically meaningless, as it would entail a mixture of elements
pertaining to different subdivisions of the ensemble represented by
$\left|{\Psi^{0}}\right\rangle $; in other words, it would imply
the simultaneous use of two partitionings of the probability space
which are incommensurable. Yet the procedure of combining under one
formula the eigenvalues that correspond to different pairs of directions
is central in the derivation of Bell-type inequalities for the bipartite
singlet spin state \cite{CHSH}. For clarity in the argument, let
us translate it to the hidden-variable language as follows.

The partitioning of the probability space $\Lambda$ corresponding
to Eq. (\ref{5-10}) can be expressed as 
\begin{equation}
C_{Q}(\boldsymbol{a},\boldsymbol{b})=\sum_{k}\intop_{\Lambda_{k}}A_{k}(\boldsymbol{a},\boldsymbol{b},\lambda)\rho(\lambda)d\lambda,\label{6-2}
\end{equation}
where 
\begin{equation}
\Lambda_{k}=\Lambda_{k}(\boldsymbol{a},\boldsymbol{b},\alpha_{k},\beta_{k})\label{6-2-1}
\end{equation}
is the probability space spanned by the subensemble represented by
$\left|{\phi^{k}}\right\rangle $; 
\begin{equation}
C_{k}(\boldsymbol{a},\boldsymbol{b})=\intop_{\Lambda_{k}}\rho(\lambda)d\lambda\label{6-3}
\end{equation}
is the corresponding statistical weight, with 
\begin{equation}
\sum_{k}\intop_{\Lambda_{k}}\rho(\lambda)d\lambda=\intop_{\Lambda}\rho(\lambda)d\lambda=1\label{6-4}
\end{equation}
in agreement with (\ref{5-13}), and 
\begin{equation}
A_{k}(\boldsymbol{a},\boldsymbol{b},\lambda)=\alpha_{k}(\boldsymbol{a},\lambda)\beta{}_{k}(\boldsymbol{b},\lambda).\label{6-6}
\end{equation}
Equation (\ref{6-2-1}) expresses the fact that the partitioning depends
on the directions $\boldsymbol{a}$, $\boldsymbol{b}$ and the eigenvalues
$\alpha_{k}$, $\beta_{k}$. The individual eigenvalues $\alpha_{k}$,
$\beta_{k}$ depend of course on $\boldsymbol{a}$ and $\boldsymbol{b}$,
respectively, as indicated in Eq. (\ref{6-6}), whilst the hidden
variables $\lambda$ themselves do not; it is only the domain $\Lambda_{k}$
which is determined by the choice of $\boldsymbol{a}$ and $\boldsymbol{b}$,
for the reasons given above.

Now, the usual starting point in the derivation of Bell-type inequalities
is the correlation written in the form \cite{CHSH} 
\begin{equation}
C_{B}(\boldsymbol{a},\boldsymbol{b})=\intop_{\Lambda}\alpha(\boldsymbol{a},\lambda)\beta(\boldsymbol{b},\lambda)\rho(\lambda)d\lambda,\label{6-7}
\end{equation}
with the probability space $\Lambda$ spanned by the ensemble represented
by the entangled state vector $\left|{\Psi^{0}}\right\rangle $. All
Bell-type derivations involve products with at least three different
orientations, say $\boldsymbol{a}$, $\boldsymbol{b}$ and $\boldsymbol{b}^{\prime}$.
In the case of the CHSH inequality, which is the more widely used
Bell-type inequality, four different orientations $\boldsymbol{a},\boldsymbol{a}^{\prime},\boldsymbol{b},\boldsymbol{b}^{\prime}$
are introduced, to write, with $\alpha=\alpha(\boldsymbol{a},\lambda)$,
a.s.o., 
\begin{align}
C_{B}(\boldsymbol{a},\boldsymbol{b})+C_{B}(\boldsymbol{a},\boldsymbol{b}^{\prime})+C_{B}(\boldsymbol{a}^{\prime},\boldsymbol{b})-C_{B}(\boldsymbol{a}^{\prime},\boldsymbol{b}^{\prime})\nonumber \\
=\intop_{\Lambda}\left[\alpha\beta+\alpha\beta^{\prime}+\alpha^{\prime}\beta-\alpha^{\prime}\beta^{\prime}\right]\rho(\lambda)d\lambda\nonumber \\
=\intop_{\Lambda}\left[\alpha(\beta+\beta^{\prime})+\alpha^{\prime}(\beta-\beta^{\prime})\right]\rho(\lambda)d\lambda\leq2.\label{6-8}
\end{align}
The inequality in the last row is obtained as an algebraic exercise
when $\alpha$, $\alpha^{\prime}$, $\beta$\textbf{, }$\beta^{\prime}$
take the values $\pm1$ only. Notice that here the same probability
space $\Lambda$ is used to construct the four correlations, without
considering the need to subdivide the ensemble in different ways depending
on the pair of orientations $(\boldsymbol{a},\boldsymbol{b})$.

According to Eqs. (\ref{6-2}-\ref{6-6}), instead, one should write,
with $A_{k}=\alpha_{k}\beta{}_{k}=\alpha(\boldsymbol{a},\lambda)\beta{}_{k}(\boldsymbol{b},\lambda)$,
a.s.o., and $\rho=\rho(\lambda)$, 
\begin{align}
C_{Q}(\boldsymbol{a},\boldsymbol{b})+C_{Q}(\boldsymbol{a},\boldsymbol{b}^{\prime}) & +C_{Q}(\boldsymbol{a}^{\prime},\boldsymbol{b})-C_{Q}(\boldsymbol{a}^{\prime},\boldsymbol{b}^{\prime})\nonumber \\
=\sum_{k}\intop_{\Lambda_{k}}\alpha_{k}\beta{}_{k}\rho d\lambda & +\sum_{l}\intop_{\Lambda_{l}}\alpha_{l}\beta^{\prime}{}_{l}\rho d\lambda+\nonumber \\
+\sum_{m}\intop_{\Lambda_{m}}\alpha_{m}^{\prime}\beta{}_{m}\rho d\lambda & -\sum_{n}\intop_{\Lambda_{n}}\alpha_{n}^{\prime}\beta^{\prime}{}_{n}\rho d\lambda,\label{6-10}
\end{align}
where $\left\{ \Lambda_{k}\right\} $, $\left\{ \Lambda_{l}\right\} $,
$\left\{ \Lambda_{m}\right\} $, $\left\{ \Lambda_{n}\right\} $ represent
the four different partitionings of $\Lambda$ corresponding to the
four different pairs of vectors. Consequently, one cannot group the
four terms under the same integral sign, as is done in passing from
the first to the second line of Eq. (\ref{6-8}). Ergo, there is no
reason why the quantum correlation $C_{Q}(\boldsymbol{a},\boldsymbol{b})$
should obey the inequality (\ref{6-8}). 

Translated to the experimental domain, this is equivalent to saying
that the spin projections $(\alpha,\beta)$, $(\alpha,\beta^{\prime})$,
a.s.o., belong to different series of experiments. Of course the experimentalist
may choose to reset the orientation of the apparatus from $\boldsymbol{b}$
to $\boldsymbol{b}^{\prime}$ after the first event, and then back
to $\boldsymbol{b}$ after the second one... But eventually, after
a large number of measurements, the experimental correlation $C_{E}(\boldsymbol{a},\boldsymbol{b})$
will be given by the average value of the projection products $\left(\alpha\beta\right)_{\boldsymbol{ab}}$,
and $C_{E}(\boldsymbol{a},\boldsymbol{b}^{\prime})$ by the average
value of the products $\left(\alpha\beta^{\prime}\right)_{\boldsymbol{ab}^{\prime}}$;
the experimentalist does not mix the data from the two series of measurements
for the calculation of the average values. If different series of
measurements are made, for different pairs of directions $(\boldsymbol{a},\boldsymbol{b})$,
one should expect the experiment to eventually confirm the functional
dependence predicted by quantum mechanics; i. e., $C_{E}(\boldsymbol{a},\boldsymbol{b})=-\boldsymbol{a}\cdot\boldsymbol{b}$.

The outcome of our present analysis leaves no room for interpretations.
As stated in Ref. \cite{Sven} in connection with the weak values,
it must be the theory that decides what meaning to ascribe to them.
The same statement applies to operators and their eigenfunctions,
as seen here in the particular case of the bipartite singlet spin
state. The Hilbert-space formalism is a powerful and elegant way of
dealing with an ensemble characterized by a common feature or physical
parameter (in our case, the total spin zero represented by $\left|\Psi^{0}\right\rangle $),
and of subdividing this ensemble according to some additional (set
of) physical parameter(s) (in our case, the pair of spin projections
onto $\boldsymbol{a}$ and $\boldsymbol{b}$). The choice of a different
physical parameter (say a spin projection along a direction $\boldsymbol{b}^{\prime}$)
implies a different partition of the ensemble. This feature needs
to be taken into account in any probabilistic analysis of the quantum
correlations.

Our conclusions, carried out entirely within the quantum formalism,
finds a counterpart in the literature in the form of the measurement-dependence
or contextuality argument. The assumption of noncontextuality (or
so-called contextuality loophole) associated with the Bell- and CHSH
theorems has been pointed out in different ways; for early works see
Refs. \cite{PeCeBr,Kup,TAB}. More recently, it is raised anew by
an increasing number of authors (see e. g. \cite{Ade,Khrenn,Vervoort,TMNiew,Kup2}),
stressing that (1) probabilities belong to experiments and not to
objects or events per se, and (2) any probability depends at least
in principle on the context, including all detector settings of the
experiment \cite{Khrenn,Vervoort}. In other words, a hidden-variable
model suffers from a contextuality loophole if it pretends to describe
different sets of incompatible experiments using a unique probability
space and a unique joint probability distribution \cite{Khrenn,TMNiew}.
\begin{acknowledgments}
The authors acknowledge support from DGAPA-UNAM through project PAPIIT
IA101918.
\end{acknowledgments}


\begin{thebibliography}{10}
\bibitem{Footnote2} Let us briefly look at the spin product for the
bipartite singlet state from the perspective of algebraic geometry.
If $\boldsymbol{s}$ and $\boldsymbol{s}'$ are two antiparallel vectors,
\[
\boldsymbol{s}\boldsymbol{a}=\exp(i\vartheta_{a}),\quad\boldsymbol{s}'\boldsymbol{b}=-\boldsymbol{s}\boldsymbol{b}=-\exp(i\vartheta_{b}),
\]
their product takes the form \cite{Hest}
\[
\left(\boldsymbol{s}\boldsymbol{a}\right)^{\ast}\left(\boldsymbol{s}'\boldsymbol{b}\right)=-\exp\left(i\theta_{ab}\right).
\]
Products containing $\boldsymbol{s}$ and $\boldsymbol{s}'$ depend
of course in general on both vectors. However, in this case the result
does not depend on $\boldsymbol{s}$ and $\boldsymbol{s}'$ because
$\boldsymbol{s}+\boldsymbol{s}'=0$, as corresponds to the (spherically
symmetric) singlet state.
\end{thebibliography}
\end{document}